\documentclass[%
 reprint,
 amsmath,amssymb,amsfonts,
 aps,
 prapplied,
 superscriptaddress,
 showkeys,
 floatfix
]{revtex4-2}

\usepackage{graphicx}
\usepackage[colorlinks = true, allcolors = blue]{hyperref}
\usepackage{float}

\begin{document}

\title{Empirical tight-binding method for large-supercell simulations of~disordered~semiconductor~alloys}

\author{Anh-Luan \surname{Phan}}
\affiliation{%
 Department of Electronic Engineering, University of Rome Tor Vergata, Via del Politecnico 1, 00133 Rome, Italy\\
}%

\author{Alessandro \surname{Pecchia}}
\affiliation{
 CNR-ISMN, Area della Ricerca di Roma 1, 00010 Rome, Italy\\
}%

\author{Alessia \surname{Di Vito}}%
\affiliation{%
 Department of Electronic Engineering, University of Rome Tor Vergata, Via del Politecnico 1, 00133 Rome, Italy\\
}%

\author{Matthias \surname{Auf~der~Maur}}
 \email{auf.der.maur@ing.uniroma2.it}
\affiliation{%
 Department of Electronic Engineering, University of Rome Tor Vergata, Via del Politecnico 1, 00133 Rome, Italy\\
}%
\date{\today}

\begin{abstract}
We analyze and present applications of a recently proposed empirical tight-binding scheme for investigating the effects of alloy disorder on various electronic and optical properties of semiconductor alloys, such as the band gap variation, the localization of charge carriers, and the optical transitions. The results for a typical antimony-containing III-V alloy, GaAsSb, show that the new scheme greatly improves the accuracy in reproducing the experimental alloy band gaps compared to other widely used schemes. The atomistic nature of the empirical tight-binding approach paired with a reliable parameterization enables more detailed physical insights into the effects of disorder in alloyed materials.
\end{abstract}

\maketitle


\section{Introduction}

Fine-tuning the electronic structure to achieve specific desired properties is one of the pivotal strategies in optoelectronic device design.
In particular, the combination of heterostructures with alloying of different materials opens up a large design space.
For example, quantum well (QW), quantum dot (QD) or superlattice (SL) structures made from pure materials or their compositional alloys have been used intensively to tailor energy levels, band gaps, or transition energies and optical strengths in order to control electronic transport, emission or absorption spectra \cite{Kamat2008Quantum,Adams2011StrainedLayer,Ekins-Daukes2013Controlling,Fox2017Quantum,Adachi2017IIIV,Alshahrani2022Emerging}.
In fact, in many cases individual layers consist of alloyed materials, such as the quantum wells in InGaN/GaN LEDs \cite{AufDerMaur2016Efficiency,Haller2018GaN}, which allows one to tune different device performance parameters by adjusting both the QW thickness and the alloy composition.
This band engineering approach has been widely applied in light emitting devices (LEDs) \cite{Ponce1997Nitridebased, Dupuis2008History}, visible and infrared detectors \cite{Pan2009Quaternary,Sun2010SiGe,Razeghi2014Advances}, and in more complex structures like quantum cascade lasers \cite{Gmachl2001Recent,Revin2011InPBased}.
Furthermore, alloying is utilized not only in material systems with well-established growth technology such as Si / Ge, III-V, III-nitrides and II-VI semiconductors, but is also applied in other systems like hybrid perovskites \cite{Rajagopal2019Understanding}.

Apparently, beside the details of QW, QD or SL structures at the device scale, the microscopic, atomistic structure in the alloyed materials has an important impact on the properties of optoelectronic devices.
Moreover, alloy-like configurations can occur at hetero-interfaces due to atomic inter-diffusion, intermixing and segregation \cite{Devaraju2012Ion,Li2017Atomic}.
Since alloys exhibit an intrinsic random disorder on the atomic scale, the local electronic and optical properties also vary spatially, to an extent that depends on the degree of uniformity and the type of specific material. Consequently, disorder can manifest itself in the macroscopic behavior of the devices, even in the ideal case of a random alloy, where the probability distribution is spatially uniform.
In other words, the versatility in material and device designs comes with the unavoidable alloy disorder, which might not be fully controllable in the technological growth processes and which can negatively impact device performance due, e.g., to the broadening of optical transitions \cite{Usman2018Largescale,Usman2018Impact} or additional alloy scattering \cite{Campman1996Interface}.
Therefore, understanding the effects of alloy disorder on electronic and optical properties is of crucial importance \cite{Mascarenhas2002Physics,Baranowski2016Review,Baranovskii2022Energy}.
From the theoretical side, this is translated into the need of models able to properly integrate disorder effects at the atomistic scale into the calculation of the electronic and optical properties of the relevant portion in the active region of a device, like a QW, a QD or a SL.

It should be noted that there are indirect methods investigating the essential aspects of electron states in disordered structures without solving explicitly the Schr\"{o}dinger equation, such as the localization landscape theory \cite{Filoche2017Localization,Piccardo2017Localization,Li2017Localization} or the low-pass filter approach \cite{Gebhard2023Quantum,Nenashev2023Quantum}. However, since the energy band structure is one of our interests here, we focus on direct band-structure calculation methods.
In this context, an atomistic model is naturally preferable to continuous-media models and effective medium approximations. One may first think about the use of \textit{ab initio} approaches like density functional theory (DFT), which can arguably provide high-accuracy results. However, being extremely computationally expensive prevents DFT from practical applications for simulating the random alloyed systems on an average computing facility. The reason is that an ideal atomistic simulations would require the consideration of very large supercells (to mitigate the impact of artificial periodicity) in multiple different atomistic configurations (to be statistically meaningful). Some authors used DFT with small supercells, especially employing the so-called special quasirandom supercells, in which one has to carefully choose a small cell configuration that best mimics the actual large supercells \cite{Wei1989Band,Zunger1990Special,Wei1990Electronic}. However, if nonuniformity or clustering is involved, or in case of QW or QD structures, then the requirement of large-supercell simulations seems to be unavoidable.

This is where the empirical tight-binding (ETB) method comes into place as a good balance between computational expense and physical accuracy.
The ETB framework set by the seminal work of Slater and Koster in 1954 \cite{Slater1954Simplified} was the father of various ever-proposed ETB schemes.
Among them, one of the most widely used in the literature is arguably the one proposed by Jancu et al. in 1998 \cite{Jancu1998Empirical} (hereafter will be referred to as ``Jancu scheme") and later given some improvements/modifications in \cite{Jancu2007Tetragonal, Zielinski2012Including,Raouafi2016Intrinsic,Nestoklon2016Virtual}.
The paper \cite{Jancu1998Empirical} supplied the ETB parameters for many IV and III-V semiconductors in an orthogonal, $sp3d5s*$, first-nearest-neighbor scheme.

During the last decades, Jancu scheme has proven its usability in various theoretical studies. In terms of disordered random alloy simulations, this scheme has been successfully applied to investigate the effect of random fluctuation in the InGaN alloy in \cite{AufDerMaur2016Efficiency, DiVito2020Simulating,Chaudhuri2021Multiscale}, for example. However, the scheme has some theoretical limitations that can cause issues when applied to alloyed systems. Motivated by that fact, in this study we will analyze these disadvantages and then mention a more recent and sophisticated ETB scheme proposed by Tan et al. \cite{Tan2016Transferable} (``Tan scheme" in the following). It will be shown that the new scheme can actually take advances in the strain treatment and the parameter-fitting procedure to reproduce to an excellent extent the experimental band gap of different alloys, also where the Jancu scheme fails.
We then demonstrate the applicability of Tan scheme in large-supercell simulations to study the impacts of alloy fluctuations on the band gap, the localization of charge carriers as well as the optical transitions for a typical antimony-containing alloy, GaAsSb. It turns out that the ETB simulation results are not only aligned well with the experimental observations, but also suggest more physical insights from the view of atomistic scale. Note that although here we select GaAsSb for detailed consideration, the workflow is clearly not limited to this specific alloy.

The paper is structured as follows. First, we will go through a brief summary of the theoretical background of the ETB method in Section \ref{sec:theory}. Next, in Section \ref{sec:schemes} the limitations of Jancu scheme and how Tan scheme overcomes those will be analyzed in detail and validated by tests to reproduce the concentration dependence of some alloy band gaps. Then, using Tan scheme, we will show how the random fluctuation effects can be investigated in the ETB framework regarding three aspects: alloy band gap, carrier localization and optical transitions. Each aspect will be discussed in its own subsection in Section \ref{sec:nonuniform}. Finally, Section \ref{sec:conclusion} is devoted to conclusions and perspectives.

\section{Theoretical background of ETB method}\label{sec:theory}
The central idea of the tight-binding method is that the single-electron wavefunction $|\psi \rangle$ of an atomistic structure can be approximately expanded in terms of a set of the tightly-bound atomic-like orbitals $\{|\phi_{\alpha,i} \rangle\}$  associated with the atoms in the unit cell, where $i$ is the atom index and $\alpha$ is the combined index for quantum numbers of the orbital, including spin. If we assume orthogonality of the basis set, then the single-electron Hamiltonian can be written as:
\begin{eqnarray}\label{eq:H}
    H &=& \sum_{(\alpha,i)} |\phi_{\alpha,i} \rangle E_{\alpha,i} \langle \phi_{\alpha,i}| ~~ +\nonumber \\
      &+& \sum_{\substack{(\alpha,i),(\beta,j)\\ (\alpha,i) \neq (\beta,j)}} |\phi_{\alpha,i} \rangle\ V_{(\alpha,i)}^{(\beta,j)} (\vec{d}_{ij}) \langle \phi_{\beta,j}|.
\end{eqnarray}
Here, the integral $E_{\alpha,i}$ and $V_{(\alpha,i)}^{(\beta,j)}$ in principle can be computed from the basis set. In fact, however, this is a very difficult and tedious task \cite{Slater1954Simplified} that we try to avoid in practical simulations. A more pragmatic way is to forget about the explicit wavefunctions of the basis orbitals and treat $E_{\alpha,i}$, $V_{(\alpha,i)}^{(\beta,j)}$ as \textit{empirical} values that can be fitted to some high-accuracy reference targets from experiments or \textit{ab initio} calculations, hence the name of the method. In the framework of ETB, $E_{\alpha,i}$ are called on-site parameters, while $V_{(\alpha,i)}^{(\beta,j)}$ are the off-site (or interatomic coupling, or hopping) parameters if $i \neq j$ and should be dependent on the magnitude and orientation of the vector $\vec{d}_{ij}$ connecting the two atoms. In the case $i=j$, $V_{(\alpha,i)}^{(\beta,i)}$ represents the intraatomic couplings between different orbitals of the same atom, which are often ignored. Based on the work of Slater and Koster \cite{Slater1954Simplified}, a numerous amount of different ETB schemes (some of them will be mentioned in the next section) have been proposed in the literature, accompanied by their corresponding sets of parameters. The main differences from one scheme to another lie in the choices of the basis set (from $sp3$ to $sp3d5s^*$), the orthogonality of these basis states, and in the number of the nearest-neighbor shells taken into account. A comprehensive topical review of ETB can be found elsewhere, e.g. in \cite{DiCarlo2003Microscopic} and the references therein.
Among others, Jancu scheme \cite{Jancu1998Empirical}, which uses orthogonal $sp3d5s^*$ basis set up to first nearest neighbor couplings, is probably the most popular for diamond and zincblende semiconductors (an extension of this scheme for nitride-containing binaries in both zincblende and wurtzite phases was given in \cite{Jancu2002Transferable}). Although this scheme or its variants has been used intensively over the past decades, in the following section we will analyze the theoretical limitations of this scheme which pose some issues when applied to certain disordered semiconductor alloys.

\section{Jancu scheme and its alloy-band-gap issue}\label{sec:schemes}

\subsection{Limitations of Jancu scheme}
The treatment of strain is certainly a concern when simulating atomistic systems with the ETB method, especially in systems with complicated and irregular strains, such as disordered random alloys. The presence of strain can lead to some or all of the following consequences:
\begin{enumerate}
    \item The change in orientation of bonds: this is naturally incorporated with the direction cosines for any ETB scheme based on Slater-Koster framework \cite{Slater1954Simplified};
    \item The change in bond lengths: this is usually taken into account by means of some distance-scaling law, which can be the generalized version of Harrison's law \cite{Harrison1999Elementary};
    \item The onsite-energy splittings of the otherwise degenerate orbitals due to strain-induced symmetry breaking;
    \item The shifts of onsite energies due to the variation of the local potential around each ions (also see Appendix \ref{app:VBO} for a discussion of the band offset). 
    \item The change of intracouplings between orbitals of the same ions due to symmetry breaking;
    \item The modification of the intercoupling (hopping) parameters;
    \item The renormalization of L\"{o}wdin orbitals \cite{Lowdin1950Non} for schemes using orthogonal basis.
\end{enumerate}

Apart from the first two effects, consideration of the others in the original Jancu scheme \cite{Jancu1998Empirical} was quite limited. For onsite splittings, the authors took only the case of uniaxial $[001]$ strain into account with a strain-induced splitting parameter $b_d$ for $d$ orbitals, but the use of macroscopic strain tensors gives rise to some ambiguities \cite{Zielinski2012Including}. The shifts of onsite energies were totally omitted, keeping the onsite and band offset parameters independent of the surrounding environment. The intracouplings were also ignored, and there are no corrections for the intercoupling parameters beside the generalized Harrison's law. Later, some modifications and improvements were proposed in \cite{Jancu2007Tetragonal,Zielinski2012Including,Raouafi2016Intrinsic,Nestoklon2016Virtual}, addressing some of the above limitations.
The approximation of the orthogonal basis can lead to severe errors in highly strained crystals or bonds. Some Slater-Koster-based ETB schemes addressed this issue by introducing on-site corrections \cite{Boykin2002Diagonal, Boykin2010Straininduced, Niquet2009Onsite} induced by strain, e.g. by performing a low-order L\"{o}wdin  rotation, but are still lacking a full consideration of all of the above effects.

In addition, there is another fatal conceptual problem in all the above-mentioned schemes and similar ones when applied to alloyed systems.
These schemes supplied ETB parameter sets for bulk pure semiconductor materials, which were fitted to the DFT calculations of the band structures for the corresponding pure materials.
The implicit assumption is that the system should be well defined in terms of pure materials. Due to this implicit assumption, the onsite energy, as well as the spin-orbit coupling (SOC) and band offset parameters of a certain ion type may differ from one pure material to another,  giving rise to the inherent ambiguity in the choice of proper parameters in alloyed systems.
A temporary workaround could be to average the onsite parameters of the component materials for each ion, according to the occurrences of the first nearest neighbors of that ion, as widely used in \cite{Li1991Electronic,Li1992Electronic,OReilly2002Tightbinding,Boykin2007Approximate,Finn2022Impact}.
For example, assuming a specific Ga cation in the GaAsSb alloy crystal has bonds with $n$ As anions and $(4-n)$ Sb anions, we then can obtain the Jancu onsite parameters (as well as SOC and band offset) for that specific Ga cation in a linear way by
$$E_{\text{Ga,GaAsSb}} = [ n E_{\text{Ga,GaAs}} + (4-n) E_{\text{Ga,GaSb}} ] /4.$$

Such a linear onsite mixing workaround is indeed physically acceptable to some extent if we make some rude approximations (see Appendix \ref{app:onsite_mixing}).
However, the validity of applying this mixing to the shear parameter $b_d$ in \cite{Jancu1998Empirical} as proposed in \cite{Zielinski2012Including} is questionable.
Another option was proposed by Carmesin et al. in \cite{Carmesin2017Interplay} to average also the intercoupling matrix elements, in addition to the above onsite mixing, within a tetrahedron and was claimed to provide better agreements for both bulk materials and nanostructures although its physical justification was not discussed.
In summary, when applied to alloy-like systems the ETB schemes relying on the concept of ``pure material" would unavoidably demand the use of certain parameter interpolation, and the choice among one interpolation or the others is not well guided by physical reasoning.

\subsection{Beyond Jancu scheme}
A more recently proposed scheme by Tan et al. \cite{Tan2016Transferable} in 2016 can be a step forward to overcome the above limitations of Jancu scheme and its variants. Although also based on the Slater-Koster framework with orthogonal $sp3d5s^*$ basis set up to the first-nearest neighbors like Jancu scheme, Tan scheme waives the concept of ``pure material".
That said, the building blocks of any atomistic structure are no longer the pure materials, but the ion species and the bonds between them. The analysis of Tan et al. is based on the idea of expanding the local net atomic potential in terms of spherical harmonics, i.e. the multipole expansion.
This allowed them to separate the contributions of different multipole components originating from the atomic potentials of the neighboring ions. In more detail, the onsite energies (and similarly, the band offset and SO couplings) of an ion are no longer fixed values but rather adopt the corrections from its (first nearest) neighbors, being adjusted by some exponential distance-scaling factors to reproduce the shifts of onsite energies.
In this way, Tan scheme saves us from choosing among ambiguous interpolations for alloy-like systems as in the ETB schemes relying on the ``pure material" concept, thus significantly improving transferability.
In fact, the Tan scheme implicitly covers the linear onsite-mixing workaround, but in a more exquisite way. For other strain effects, the Tan scheme features the intracouplings induced by the dipole and quadrupole components of the local net potential whenever the symmetry of the crystal is reduced, which at the same time implies the possible splitting of the degenerate orbitals. Moreover, the multipole components give corrections to the intercouplings between two ions, taking into account also the information about the positions of the neighboring ions of the pair. Note that this is another advance of the Tan scheme when it partly considers the three-center terms which are omitted in the two-center approximation of the Slater-Koster framework \cite{Slater1954Simplified}. All of these amendments allow the Tan scheme to naturally deal with, as they claimed, arbitrary strain profiles, at the price of a much larger number of fitting parameters than any ETB scheme previously proposed. The abilities of dealing with arbitrary strain profiles and retaining to a large extent the information of chemical species are crucial points in encountering with the disordered nature of random alloys.

For any empirically fitting method, the transferability is an important factor of merit. Apart from the sophistication level of the model itself, the transferability depends (perhaps even more critically \cite{Zielinski2012Including}) on the quality of the parameter fitting procedure \cite{Tan2013Empirical,Tan2015Tightbinding}. In this regard, while most of the other schemes considered the features of the band structures as the only fitting targets, Tan et. al. used the fitting process from \cite{Tan2015Tightbinding}, which involves both the DFT band structures and the wave functions. With extra information for fitting, one can reasonably expect higher transferability for parameter sets of the Tan scheme, especially when one works with the optical simulations of the materials in which the appropriate wavefunctions of the charge carriers are needed to calculate the optical transitions. The last but not least reason for our choice of Tan scheme instead of the other alternatives is the availability of the ETB parameter sets for various IV and III-V materials.
The main downside of this scheme is the inflation in the number of parameters, but it will only concern those who want to parameterize the new materials, not those who use the scheme with already established parameters.

\subsection{The alloy band-gap problem} \label{subsec:gap_problem}
To validate the above discussion about the two schemes, we computed and compared the band gap variation of three Sb-containing alloys GaAsSb, InAsSb, InGaSb in terms of the Sb- or In-concentration.
To a good approximation, the concentration-dependence of a ternary alloy's band gap usually follows a quadratic function \cite{Adachi2017IIIV}, written as
\begin{eqnarray} \label{eq:gap_bowing}
    E^g_{\text{alloy}} = x E^g_{\text{A}} + (1-x) E^g_{\text{B}} - b x (1-x),
\end{eqnarray}
where $x$ is the concentration of the alloy $\mathrm{A}_x\mathrm{B}_{1-x}$ composed of binary constituents $\mathrm{A}$ and $\mathrm{B}$, and the coefficient $b$ is the so-called bowing parameter. The estimated values of $b$ from the experimental data for various III-V ternary alloys can be found in the literature, e.g. in \cite{Vurgaftman2001Band}. In the following comparison, we will assess how well Jancu scheme (with onsite-mixing workaround) and Tan scheme can reproduce the experimental direct-gap bowing of the above three alloys. For all simulations in this work, we use the multiscale optoelectronic simulation software TiberCAD \cite{AufDerMaur2007TiberCAD,AufDerMaur2008TiberCAD}, in which we have implemented both ETB schemes. We chose large cubic supercells of 12 $\times$ 12 $\times$ 12 nm$^3$ to diminish the error due to the use of artificial periodicity. Furthermore, to ensure the statistical significance of the results, we performed simulations for $30$ different supercell configurations for each concentration being considered, and took the configurational averages afterward. Some other minor details of the simulation setup are described in Appendix \ref{app:setup}.

\begin{figure}[htb!]
    \begin{minipage}[b]{\linewidth}
    \includegraphics[width=\linewidth]{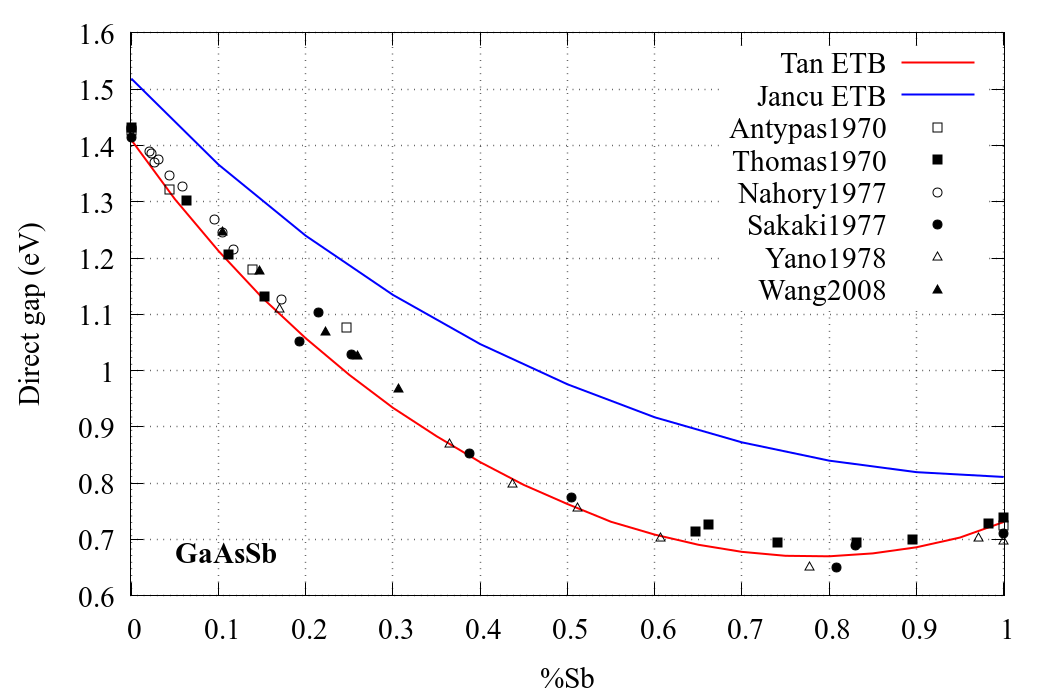}
    \end{minipage}
\hfill
    \begin{minipage}[b]{\linewidth}
    \includegraphics[width=\linewidth]{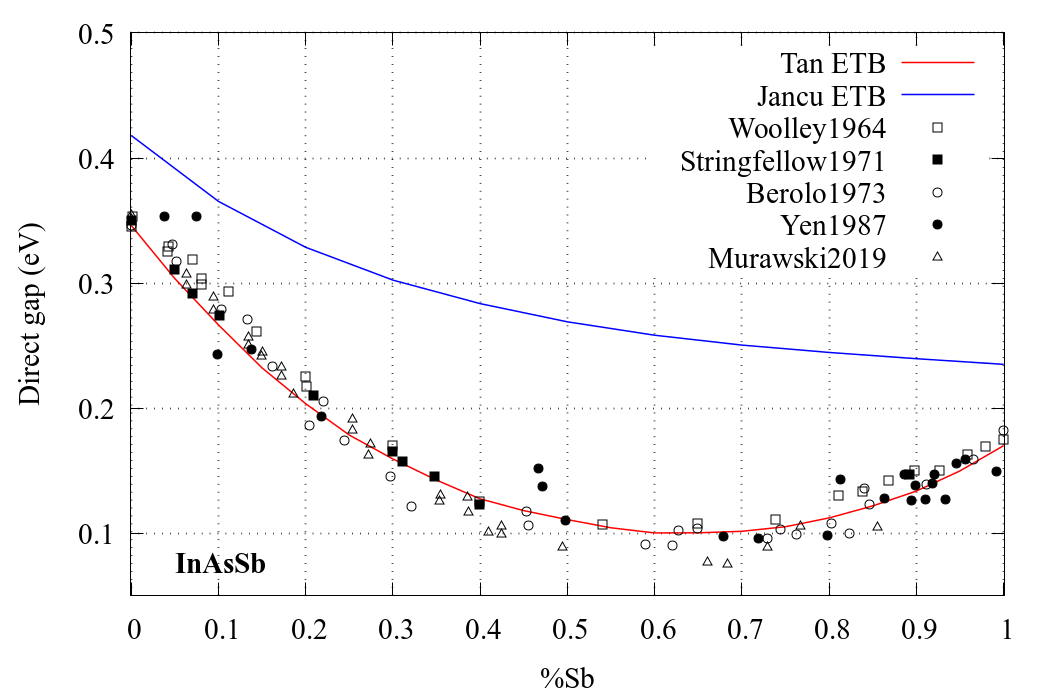}
    \end{minipage}
\hfill
    \begin{minipage}[b]{\linewidth}
    \includegraphics[width=\linewidth]{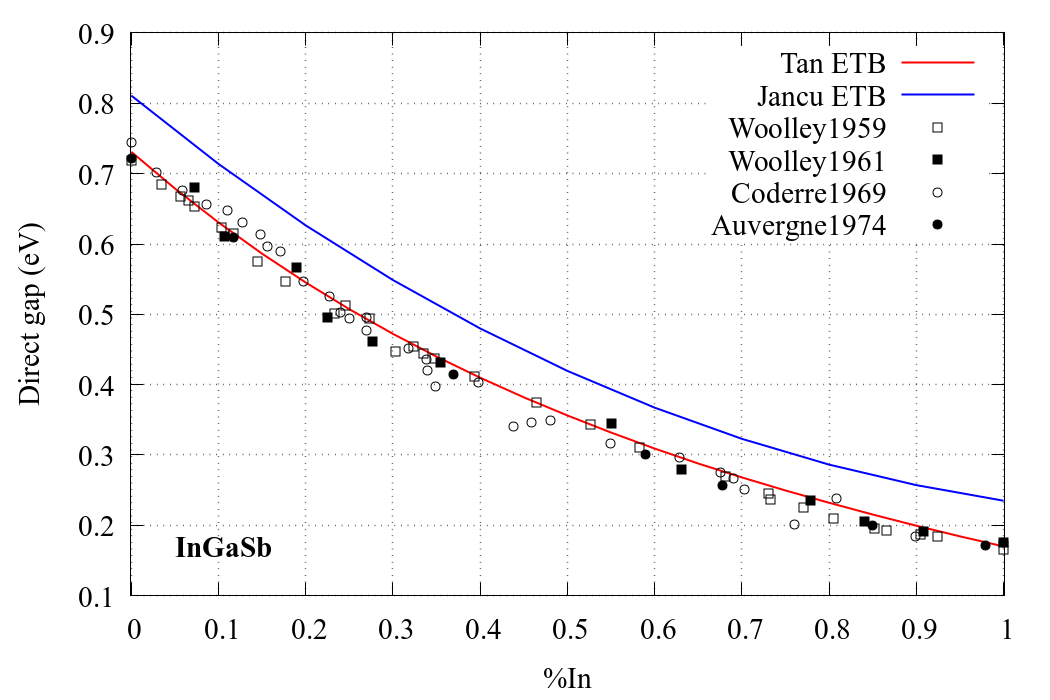}
    \end{minipage}
\caption{\label{fig:bowing} Variations in band gaps versus the concentration for totally random GaAsSb, InAsSb and InGaSb alloys calculated by Tan scheme (300K) and Jancu scheme (0K). Experimental data (300K) from various sources are shown for comparison purposes: Woolley1959 \cite{Woolley1959Optical}, Woolley1961 \cite{Woolley1961Temperature}, Coderre1969 \cite{Coderre1969Conduction}, Auvergne1974 \cite{Auvergne1974Piezoreflectance}; Woolley1964 \cite{Woolley1964Optical}, Stringfellow1971 \cite{Stringfellow1971Liquid}, Berolo1973 \cite{Berolo1973Effect}, Yen1987 \cite{Yen1987Molecular}, Murawski2019 \cite{Murawski2019Bandgap}; Antypas1970 \cite{Antypas1970Liquid}, Thomas1970 \cite{Thomas1970Energy}, Nahory1977 \cite{Nahory1977Growth}, Sakaki1977 \cite{Sakaki1977InGaAs}, Yano1978 \cite{Yano1978Molecular}, Wang2008 \cite{Wang2008Characterization}.}
\end{figure}

The results are presented in Fig. \ref{fig:bowing}. Note that the differences in the band gaps of the pure materials are due to the fact that Jancu ETB parameters were fitted to $0$K targets, while those of Tan scheme were fitted to $300$K targets. For better comparison with the experiments, we also plot the experimental band-gap values at $300$K from various references (see the caption).
It is interesting that Tan scheme reproduces excellently the band-gap bowing of all these alloys, following most of the experimental data points. In contrast, the Jancu scheme works well only for the case of the common-anion alloy InGaSb with small gap-bowing and significantly underestimates the large bowing parameters in the case of common-cation alloys GaAsSb and InAsSb. As a result, the Jancu scheme predicts no gap minima for GaAsSb and InAsSb alloys while they indeed exist. It should be emphasized that the existence of a gap minimum is one of the interesting features of alloying because it extends the tunable energy range for optoelectronic applications. These results validate our discussions in the previous subsections about the advantages of Tan scheme over Jancu scheme for the large-supercell simulations of the disordered semiconductor alloys, since correctly reproducing the experimental band gap of an alloy clearly is of major importance in modeling disordered alloys or superlattices.

\section{Effects of random fluctuations in G\lowercase{a}A\lowercase{s}S\lowercase{b}} \label{sec:nonuniform}
So far we have assessed the applicability of the ETB method for the cases of uniform alloys, in which the parent materials of the alloys are distributed over the supercell with a uniform probability. This ideal situation is not necessarily the case in reality. In fact, the distribution of the parent materials can be in a nonuniform manner, favoring the formation of clusters. This alloy fluctuation may bring more complexities to the electronic and optical properties of the alloy samples. In what follows, we will evaluate these fluctuation effects on the GaAsSb alloy using the atomistic ETB simulations in the TiberCAD software and compare the simulation results with the reported experimental observations. 
GaAsSb alloy is one of the most important compounds of the III-V-based semiconductor material group thanks to its unique properties, which are promising for potential applications in telecommunication and optoelectronics \cite{Anabestani2021Review}. However, the introduction of Sb into GaAs to form GaAsSb can introduce compositional fluctuation and in turn leads to localized states, which will deteriorate the useful electronic and optical properties of the alloy as experimentally confirmed in \cite{Gao2016Investigation}. As shown in the following, our atomistic ETB simulations can give good agreements with \cite{Gao2016Investigation}. Although we chose GaAsSb for a detailed investigation, the workflow here is not limited to this specific alloy.

In order to take spatial alloy nonuniformity into account, in addition to the simulation setup described in Section \ref{subsec:gap_problem}, we also control how uniformly Sb or As will be distributed in the anion sublattice.
For example, if the Sb percentage is less than 50\% (As-rich regime) then we let only a certain percentage (which we call uniformity level) of the minority anion type (i.e. Sb) be distributed with a uniform probability across the supercell.
The remaining minority anions will have a higher probability of being placed close to the already present minority anions.
The roles of the two anion types are interchanged if the Sb concentration is above 50\% (Sb-rich regime). For the 50\% alloy, we performed calculations for both possibilities.
A uniformity level of $100\%$ is equivalent to the totally random case that has been considered in Subsection \ref{subsec:gap_problem}, while a smaller uniformity level indicates a higher possibility of cluster formation.

\subsection{Impacts on alloy band gap}
\begin{figure}[htb!]
    \begin{minipage}[b]{\linewidth}
    \includegraphics[width=\linewidth]{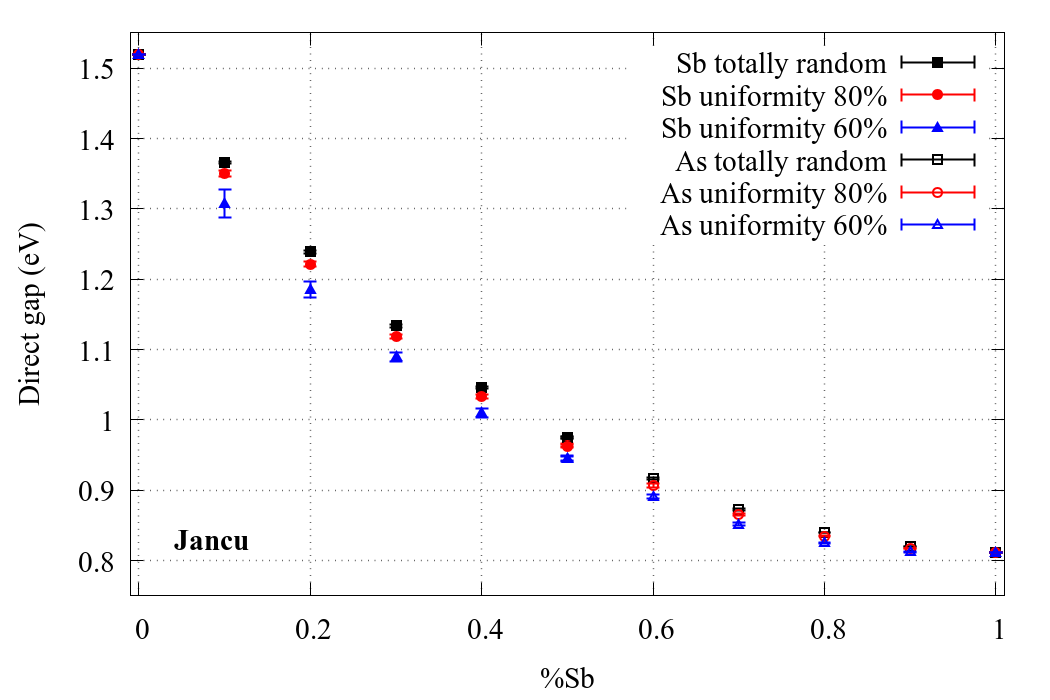}
    \end{minipage}
\hfill
    \begin{minipage}[b]{\linewidth}
    \includegraphics[width=\linewidth]{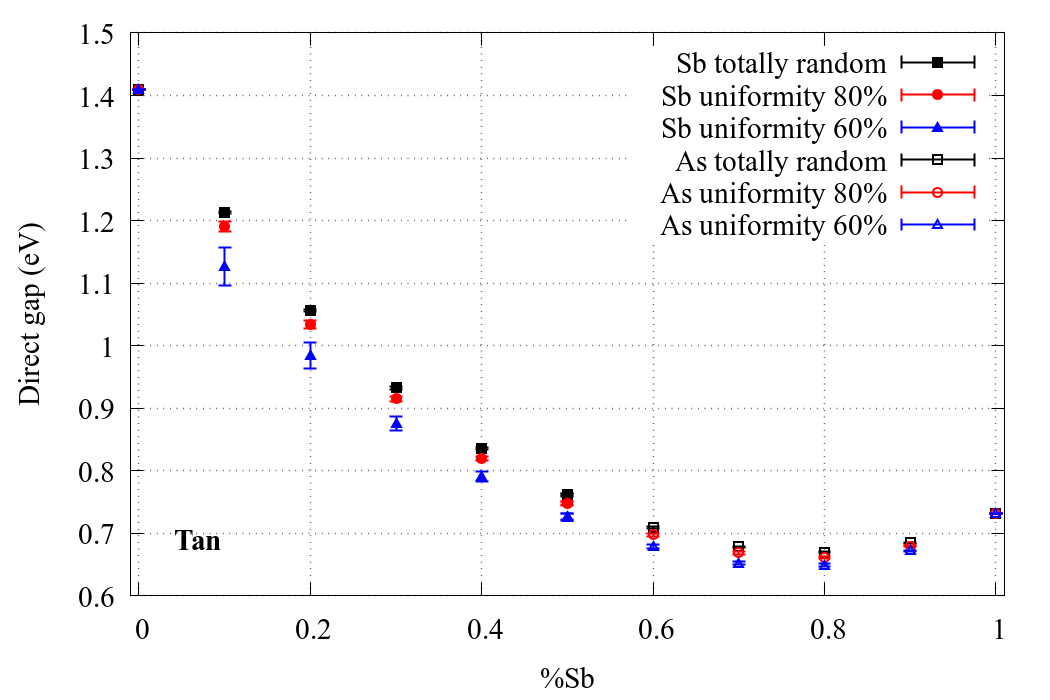}
    \end{minipage}
\caption{\label{fig:nonuniform} The variation of the GaAsSb band gap versus Sb-concentration but for different levels of anion uniformity calculated by the Jancu scheme (at 0K, upper panel) and the Tan scheme (at 300K, lower panel) represented by mean values and statistical error bars. Depending on whether the alloy is As-rich ($\% \text{Sb} \le 0.5$) or Sb-rich ($\% \text{Sb} \ge 0.5$), we choose the minority anion type to control the uniformity level. The Sb (As) uniformity means the percentage of Sb (As) anions that are uniformly distributed throughout the supercell.}
\end{figure}

In Fig. \ref{fig:nonuniform} we show the mean values as well as the statistical scatterings of the GaAsSb band gap with respect to the Sb concentration for different uniformity levels obtained by both Jancu scheme and Tan scheme. We can see that apart from the gap bowing, the two schemes agree qualitatively with each other. The results show that the incorporation of just a small amount of Sb into GaAs causes the band gap to decrease dramatically. The nonuniform distribution of these Sb ions in the material decreases the band gap even further, while also increasing the statistical scattering of the value.
The bowing of band gap variation thus can be increased locally in the small range of Sb concentration by the nonuniform Sb distribution compared to the expected value in the ideal uniform situation, which is a noteworthy factor for gap bowing in experimental measurements. However, for increasing Sb content, the difference between the uniform and nonuniform cases gradually decreases.

Interestingly, the impact of nonuniformity is not symmetric between the Sb nonuniformity in As-rich regime (the left side of Fig. \ref{fig:nonuniform}) and As nonuniformity in Sb-rich regime (the right side). Namely, the nonuniform distribution of As in the Sb-rich regime leads to only a negligible decrease in the band gap when the As concentration is small, and the impact increases just a bit for higher As concentration. This suggests that experimentally, the formation of As-rich regions has less impact on material properties than the formation of Sb-rich clusters, and that the latter should be more easily identifiable by e.g. optical measurement. On the other hand, this may give us another possible degree of freedom in tuning the band gap of alloys.

\subsection{Impacts on carrier localization}\label{subsec:localization}
\begin{figure}[htb!]
    \includegraphics[width=\linewidth]{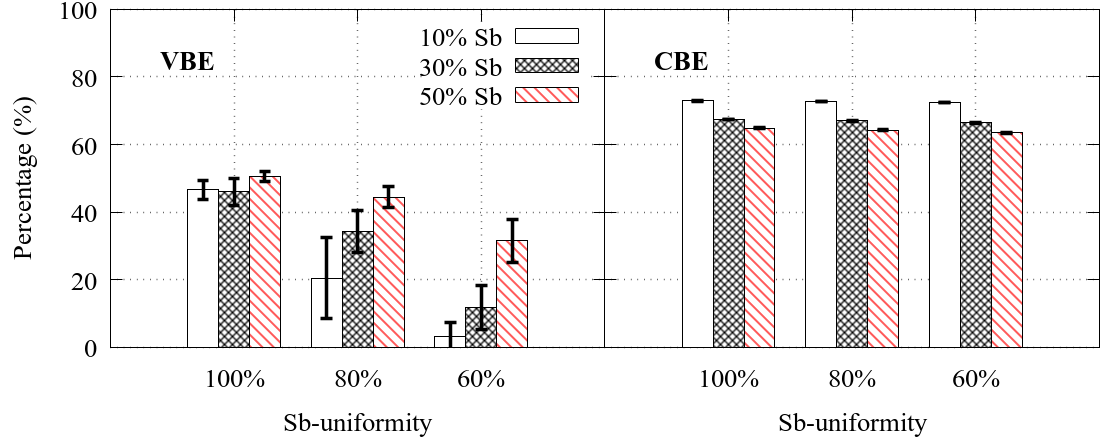}
    \caption{The percentage of top contributing atoms to $80\%$ density of the VBE (left) and CBE (right) wavefunctions versus the Sb-uniformity level in As-rich regime for $\% \text{Sb}$ from $0.1$ to $0.5$.}
    \label{fig:contributorSb}
\end{figure}

\begin{figure}[htb!]
    \includegraphics[width=\linewidth]{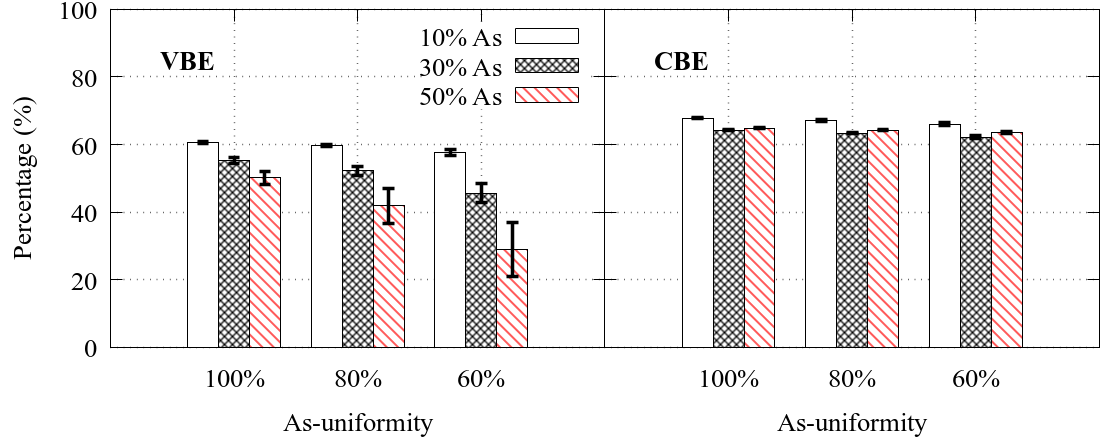}
    \caption{Similar to Fig.~\ref{fig:contributorSb} but versus As-uniformity in Sb-rich regime for $\% \text{Sb}$ from $0.5$ to $0.9$ (equivalen to \%As from 0.5 down to 0.1).}
    \label{fig:contributorAs}
\end{figure}

To gain insight into how alloy fluctuations affect the carrier localization, we calculated the projected density of states (PDOS) of each ion in the supercell and then extracted how many of them contribute the most to $80\%$ of the probability density of states at the valence-band edge (VBE) and the conduction-band edge (CBE). A smaller percentage implies a stronger localization of the charge carrier. These results are demonstrated in Figs. \ref{fig:contributorSb} and \ref{fig:contributorAs} for Tan scheme only. Results obtained with Jancu scheme are similar.
From Figs. \ref{fig:contributorSb} and \ref{fig:contributorAs} it is clear that the anion uniformity, regardless of Sb or As, in general makes the VBE state (hole) more localized and also increases the statistical scattering. On the other hand, nonuniformity has a negligible impact on the localization of the CBE state (electron). Our ETB simulation results support the experimental observations of Gao et al. \cite{Gao2016Investigation} about the existence of localized states in the growth of GaAsSb alloy, which was attributed to the fluctuations of the Sb distribution in the random alloy. It should be noted that in \cite{Gao2016Investigation} all the measured GaAsSb samples were in the As-rich regime with $\%\text{Sb} < 10\%$. Here, with the ETB simulations spanning a full range of concentrations, we can draw a more general conclusion that the nonuniformity of the anion distribution, not only of the Sb distribution, enhances the carrier localization in the GaAsSb alloy. Moreover, we point out in particular that the ones that get more localized are the hole-like states, while the electron-like states remain substantially delocalized.

Another notable point is that a high concentration of Sb can mitigate the impact of Sb nonuniformity in the As-rich regime, as we can observe from the left plot of Fig. \ref{fig:contributorSb} for different Sb concentrations. This seems to be contrary to \cite{Gao2016Investigation} at first glance since they concluded that the degree of localized states increases with increasing Sb content. However, note that the concentrations in their samples are below 10\%. In fact, the degree of localization should begin to increase rapidly from the pure GaAs case, then reaching a maximum at some percentage well below 50\%, before decreasing with more and more Sb added.
The opposite trend occurs for As nonuniformity in the Sb-rich regime, where a large As concentration tends to enhance the localization of the hole state. Together with the observation of the band gap variation in Fig. \ref{fig:bowing}, this shows that the impacts of Sb and As nonuniformity in the GaAsSb alloy are asymmetric.

This asymmetry can be understood qualitatively if we pay attention to the relative band offset between GaAs and GaSb. These two binary materials have a quite large valence band offset (the natural VBE of GaSb is about 0.77 eV higher than that of GaAs, according to \cite{Vurgaftman2001Band}). When nonuniformity is present, clusters of GaSb (or GaAs) may form, acting like local potential quantum wells (or barriers) with respect to the holes. In the As-rich regime, even a small amount of GaSb clusters can trap the holes within a small space, thus suddenly enhancing the hole localization compared to the pure bulk GaAs. 
On the other hand, in the Sb-rich case a locally increased As concentration induces a circumscribed potential barrier, which cannot considerably change hole localization.
For the CBE state, since the CBEs of GaAs and GaSb are almost aligned, the fluctuation of local potential is very weak and consequently has a negligible impact on the movement of lightweight electrons.
In conclusion, the asymmetry between the impact of Sb- and As-nonuniformity is in intimate relation to the asymmetry between the band offsets of GaSb and GaAs.

\subsection{Impacts on optical transitions}\label{subsec:optical}
\begin{figure}
    \includegraphics[width=\linewidth]{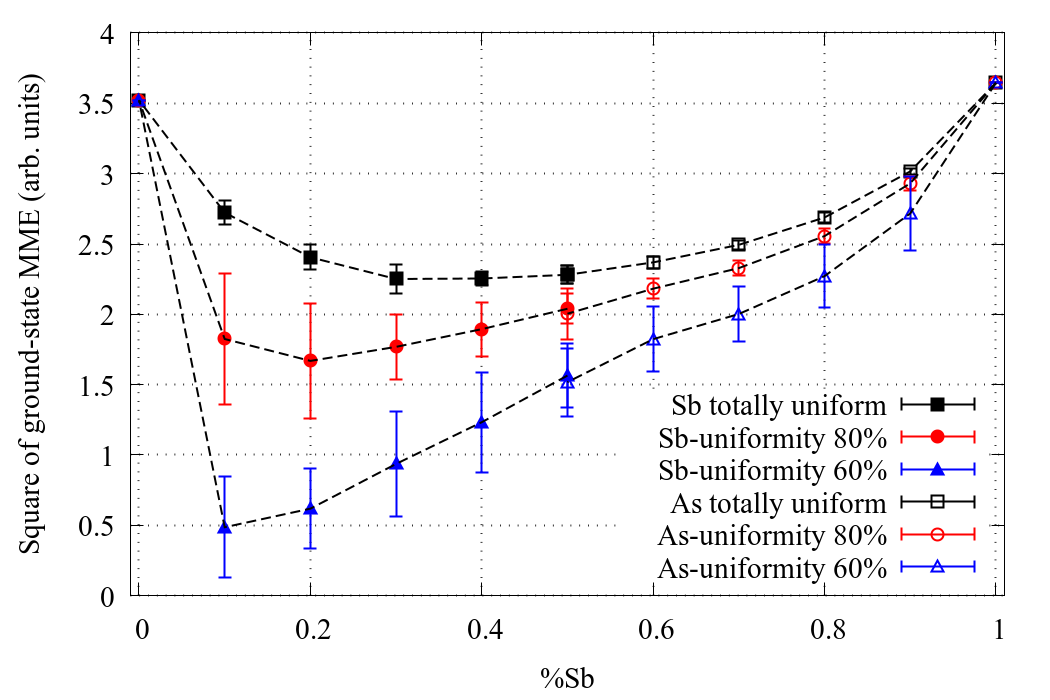}
    \caption{The square of ground-state transition momentum matrix element (MME) in atomic units at $\Gamma$ point for GaAsSb alloy with different uniformity levels of anion distribution. Black dashed lines are for eye-guiding purpose.}
    \label{fig:MME2}
\end{figure}

As discussed in Section \ref{sec:schemes}, Tan scheme advances Jancu scheme in describing the proper alloy band gap. On the other hand, the fitting procedure of Tan scheme \cite{Tan2015Tightbinding}
is expected to treat the wavefunction-depedent quantities like the optical transitions more accurately. These characteristics are important if one wants to use the ETB method for simulating optical properties of the disordered semiconductor alloys. In the following, we apply the Tan scheme to calculate the momentum matrix element (MME) of the ground-state transition at the $\Gamma$-point of the Brillouin zone of GaAsSb as a function of Sb content with different uniformity levels of the anion distribution. The result is shown in Fig. \ref{fig:MME2}. One can see that the magnitude of the MMEs decreases significantly with the increase in nonuniformity level, whereas their statistical scatterings become larger in general. This can be attributed to smaller overlapping between the wavefunctions of the electron and hole states, which is directly related to the enhanced localization of the hole states in the presence of nonuniformity. The effect is consistent with the analysis in Section \ref{subsec:localization}. Furthermore, in the As-rich regime, the scatterings in MME values increase rapidly at higher Sb concentrations, especially in the presence of nonuniformity, from which we would expect a larger broadening in the spontaneous emission spectra. This result again confirms the experimental observations in \cite{Gao2016Investigation}. Incorporating a small amount of Sb into GaAs can lead to a substantial decrease in optical transition strength if Sb ions are not distributed uniformly, which could result in a loss of efficiency in optoelectronic applications. The higher the nonuniformity, the more sensitive the MMEs are with respect to the alloy composition.

\begin{figure}
    \includegraphics[width=\linewidth]{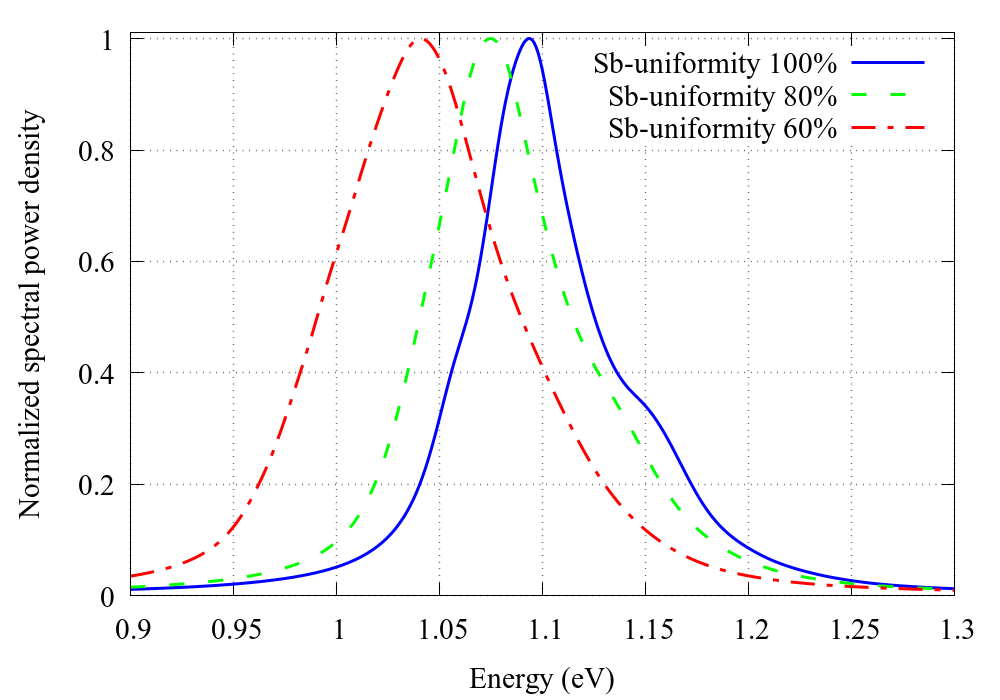}
    \caption{The normalized spontaneous emission spectra of GaAsSb for 20\% Sb with different Sb-uniformity levels at $300$K.}
    \label{fig:spectra}
\end{figure}
We also calculated the (normalized) spectral power density of GaAsSb in As-rich regime ($20\%$ Sb) for different uniformity levels at $300$K, presented in Fig. \ref{fig:spectra}, showing that apart from a red-shift due to the band gap narrowing also a spectral broadening should be expected. Finally, the effects on the optical transition in the Sb-rich regime is much less sensitive to the concentration and the uniformity level of As ions, which again manifests the asymmetric roles between the two anion types.

\section{Conclusions and outlooks} \label{sec:conclusion}
As a balance between the costly \textit{ab-initio} calculations and the continuous-media models like k$\cdot$p, the ETB method has found its applications in simulations involving large numbers of atoms where the details of atomistic arrangement matter, like in the case of random semiconductor alloys with possibly nonuniform ion distributions.
This work discussed the limitations of a widely used ETB scheme \cite{Jancu1998Empirical} in dealing with the complexities due to irregular strain and chemical species profiles in such materials. We showed that a more state-of-the-art ETB scheme \cite{Tan2016Transferable} can take an advance over these limitations and reproduce the results that are aligned well with the experimental measurements, especially for the concentration dependence of the alloy's band gaps.
We then demonstrated how the new ETB scheme can serve as a suitable tool for studying the effects of alloy disorder and nonuniform ion distributions on the electronic and optical properties of alloy materials via a specific example of GaAsSb. The calculations show that, as expected, nonuniformity further decreases the alloy's band gap, enhances in particular hole localization and degrades the strength of optical transitions. All of these ETB simulation results not only agree well with experimental observations \cite{Gao2016Investigation}, but also add more physical insights from the atomistic view.
Finally we would like to highlight that the ETB simulations can be applied transferably to investigate other alloys, QWs, QDs, SLs, heterostructures with interfaces between different materials, or in general atomistic systems of reasonable sizes, provided that the necessary ETB parameters are available.

\begin{acknowledgments}
The authors thank Dr. Yaohua Tan for the valuable discussions on his work \cite{Tan2016Transferable}.
This work is supported by European Union’s Horizon 2020 research and innovation programme under the Marie Skłodowska-Curie grant agreement nr. 956548, project Quantimony.
MAdM acknowledges support by the European Union—NextGenerationEU under the Italian National Center 1 on HPC—Spoke 6: “Multiscale Modelling and Engineering Applications” MUR CUP: E83C22003230001.
\end{acknowledgments}

\appendix

\section{The role of band offset} \label{app:VBO}
The concept of band offset used in this work refers to the natural band offset of a material in reference to a common energy level, e.g. the vacuum energy level, when it is not in contact with any other materials. This quantity is not the same as, though closely related to, the band offset at the interfaces of the heterostructures.
In the ETB framework, the natural band offset of a material is usually taken into account by the valence band offset (VBO) parameter. Plotting the VBO values of different materials shows how the valence band edges (VBE), and consequently the whole band structures, of these materials are aligned with each other. The question is why band offset (or VBO) matters in alloy simulations.
Mathematically speaking, if we shift all the onsite parameters in Eq. \ref{eq:H} of a pure material, like GaAs or GaSb, by the same amount then the band structure of that pure material is unchanged except for being rigidly shifted by the same amount, leading to an arbitrariness in choosing the onsite energy values of the ions, which in turn make the alloy calculation of GaAsSb nonsense. Thus, a physical choice should be made: firstly, one can parameterize the ``raw" onsite parameters (together with other ETB parameters) so that the VBEs of all pure materials lie at a common level, says 0 eV, and then add the corresponding VBO parameters to the ions' ``raw" onsite parameters to reproduce the relative alignment between the band structures of different materials.
The VBO parameters thus bear the information about the background electric potential that the electron experiences in different lattice environments. Therefore, the proper band offset (or VBO parameter) must be taken into account, namely added to the ``raw" onsite parameters to make ETB calculations reasonable.
This is not only true for alloy simulations, but also all kinds of ETB simulations that involve different pure materials together, e.g. superlattices and in general heterostructures. The VBOs were not considered as fitting parameters in the Jancu scheme \cite{Jancu1998Empirical} because the goal of that paper was to reproduce the results of pure bulk materials only. In contrast, Tan scheme \cite{Tan2016Transferable} explicitly includes the offset, which is a further advantage over Jancu scheme.

\section{The justification of the onsite-mixing workaround} \label{app:onsite_mixing}
Here we give a brief proof of why the onsite-mixing workaround is physically reasonable.
First, we assume that the total potential of the single-electron Hamiltonian is a sum of spherical atomic-like potentials $V_{\vec{R}}$ due to the ions at $\vec{R}$:
\begin{eqnarray*}
    H = T + \sum_{\vec{R}} V_{\vec{R}} (\vec{r}).
\end{eqnarray*}
Calculating the diagonal matrix element for an orbital of a specific ion, one can separate the contributions to that matrix element as follows: a fixed term due to the potential of the ion of interest, and then terms from its first, second, etc. nearest-neighboring shells. The latter terms are linear with respect to the electric potentials and depend on the local lattice environment around the central ion. In Jancu scheme \cite{Jancu1998Empirical}, unfortunately the diagonal matrix elements are computed from fixed orbital onsite parameters. If we insist that Jancu parameterization implicitly captures the underlying physics, it makes sense to assume the contributions from the neighboring shells in different pure materials are already embedded in the corresponding onsite parameters. Now, starting from a pure material and substituting some of the ions in the neighboring shells by another ion type to get an alloyed system, the contributions from those substituted neighboring ions will change proportionally to the magnitude of the new potentials. With some simple mathematical arrangements, it turns out, in the case of ternary alloy GaAsSb for example, the onsite parameters of a specific Ga cation in alloy $E_{\text{Ga}}^{\text{alloy}}$ can be roughly deduced from those in GaAs $E_{\text{Ga}}^{\text{GaAs}}$ and in GaSb $E_{\text{Ga}}^{\text{GaSb}}$ as
\begin{eqnarray}\label{eq:average}
E_{\text{Ga}}^{\text{alloy}} \approx \dfrac{n_{\text{As}} E_{\text{Ga}}^{\text{GaAs}} + n_{\text{Sb}} E_{\text{Ga}}^{\text{GaSb}}}{n_{\text{As}} + n_{\text{Sb}}} + \mathcal{O}(III)
\end{eqnarray}
with $n_{\text{As/Sb}}$ are the number of As/Sb anion in the first nearest shell and $\mathcal{O}(III)$ stands for the contributions from the third or further shells. To get to this Eq. \ref{eq:average}, one has to assume that the contributions from an ion type to its neighbor's onsite energies remain the same regardless of the possible changes in bond lengths and in orbital wavefunction. Omitting $\mathcal{O}(III)$ in Eq. \ref{eq:average} gives us a linearly weighted average of $E_{\text{Ga}}^{\text{GaAs}}$ and $E_{\text{Ga}}^{\text{GaSb}}$ according to the occurrences of the corresponding anion in the first shell. In short, this result is inherent from the fact the electric potentials can be linearly superposed.
Similar arguments apply for band offset and spin-orbit coupling parameters. Notably, this onsite-mixing workaround also ensures that the alloy's natural band structure is properly aligned with respect to those of pure materials.
It should be noted that, although the Jancu onsite parameters of the same ion type in different pure materials differ very little, the band offsets in contrast usually vary quite much from one material to another (e.g. see the values documented in \cite{Vurgaftman2001Band}) so that the weighted averages would change the diagonal Hamiltonian matrix elements considerably.

\section{Setup for large-supercell alloy simulations}\label{app:setup}
In the framework of TiberCAD software, the atomistic structures of the alloys are automatically generated based on a reference lattice. For example, in the case of GaAsSb, initially a normal GaAs zincblende lattice, whose lattice constant is given by Vegard's law for ternary alloys, is created filling the volume of the supercell. Next, an amount of As anions will be substituted by Sb anions according to some probability distribution depending on the alloy's concentration, the uniformity level and a chosen random-number-generating seed. The whole supercell is then relaxed using a simple valence force field (VFF) approach \cite{Camacho2010Application}, resulting in new positions of all ions. This new atomistic structure is passed to the ETB module to build the Hamiltonian matrix, which will be solved for the eigen-solutions, using either Jancu or Tan schemes. Optionally, we can calculate the projected density of state (PDOS) and the optical matrix elements. The valence band offsets were taken from \cite{Vurgaftman2001Band}.

We would like to note that there are some minor differences in our implementation of Tan scheme compared to the description in the original paper \cite{Tan2016Transferable}:
\begin{itemize}
    \item For ions in alloys, we kept the values of parameters $C$ specific for each of its bonds to the neighbors instead of using the averaged values;
\item    We swapped the values between the rows for $E_{s^*}$ and $E_d$ in their Table II, as from the calculated bandstructures they appear to have been accidentally interchanged;
\item The matrix element at row $3$, column $4$ in their Equation (A4) should be $xz/\sqrt{3}$;
\item We insisted on using the valence band offsets given in \cite{Vurgaftman2001Band} by adding some rigid shifts to the onsite parameters.
\end{itemize}

\bibliography{references}
\end{document}